\documentclass[sigconf]{acmart}
\usepackage{soul}
\usepackage{multirow}
\usepackage{epstopdf}
\usepackage{hyperref}
\usepackage{listings}
\usepackage{fancybox}
\usepackage{graphicx}
\usepackage{subcaption}
\usepackage{color}
\usepackage{paralist}
\usepackage{ragged2e}
\usepackage{enumitem}
\usepackage{algorithm}
\usepackage{algorithmic}
\usepackage{booktabs}
\usepackage[utf8]{inputenc}
\usepackage{listings}

\usepackage[framemethod=TikZ]{mdframed}
\usepackage{tcolorbox}
\makeatletter
\newcommand{\mybox}[1]{%
  \setbox0=\hbox{#1}%
  \setlength{\@tempdima}{\dimexpr\wd0+13pt}%
  \begin{tcolorbox}[boxrule=0.5pt, colback=white, arc=4pt,
      left=6pt,right=6pt,top=6pt,bottom=6pt,boxsep=0pt]
    #1
  \end{tcolorbox}
}
\usepackage{tikz}
\usepackage{color}
\usepackage{xcolor}
\usepackage{array}
\usepackage{amsmath}
\usepackage{centernot}
\usepackage{xspace}
\usepackage{url}
\usepackage{verbatim}
\usepackage{wrapfig}
\usepackage{tabularx}
\definecolor{javared}{rgb}{0.6,0,0} 
\definecolor{javagreen}{rgb}{0.25,0.5,0.35} 
\definecolor{javapurple}{rgb}{0.5,0,0.35} 
\definecolor{javadocblue}{rgb}{0.25,0.35,0.75} 

\newcommand{\ReCom}{Re\textsuperscript{2}Com~}

\AtBeginDocument{%
  \providecommand\BibTeX{{%
    \normalfont B\kern-0.5em{\scshape i\kern-0.25em b}\kern-0.8em\TeX}}}

\setcopyright{acmcopyright}
\copyrightyear{2018}
\acmYear{2018}
\acmDOI{10.1145/1122445.1122456}

\acmConference[Conference’21]{In Proceedings
of ACM Conference (Conference’21)}{June 03--05, 2018}{Woodstock, NY}
\acmBooktitle{Proceedings of ACM Conference (Conference’21),
  June 03--05, 2018, Woodstock, NY}
\acmPrice{15.00}
\acmISBN{978-1-4503-XXXX-X/18/06}

\begin{document}

\title{Yet Another Combination of IR- and Neural-based Comment Generation}

\author{Yuchao Huang}
\affiliation{%
	\institution{Chinese Academy of Sciences}
	  \city{Beijing}
	\country{China}
}
\email{yuchao2019@iscas.ac.cn}

\author{Moshi Wei}
\affiliation{%
	\institution{York University}
		 \city{Toronto}
	\country{Canada}
}
\email{moshiwei@yorku.ca}

\author{Song Wang}
\affiliation{%
	\institution{York University}
	 \city{Toronto}
	\country{Canada}
}
\email{wangsong@yorku.ca}

\author{Junjie Wang}
\affiliation{%
	\institution{Chinese Academy of Sciences}
	 \city{Beijing}
	\country{China}
}
\email{junjie@iscas.ac.cn}

\author{Qing Wang}
\affiliation{%
	\institution{Chinese Academy of Sciences}
	 \city{Beijing}
	\country{China}
}
\email{wq@iscas.ac.cn}


\begin{abstract}
\textbf{Background:} Code comment generation techniques aim to generate natural language descriptions for source code. There are two orthogonal approaches for this task, i.e., 
information retrieval (IR) based and neural-based methods.
{Recent studies have focused on combining their strengths by feeding the input code and its similar code snippets retrieved by  the IR-based approach to the 
{neural-based}
approach, which can enhance the neural-based approach’s ability to output low-frequency words and further improve the performance.}

\textbf{Aim:} 
However, despite the tremendous progress, 
our pilot study reveals that the current combination is not generalizable and can lead to performance degradation.  
In this paper, we propose a straightforward but effective approach to tackle the issue of existing combinations of these two comment generation approaches. 

\textbf{Method:} 
{Instead of binding IR- and neural-based approaches statically, we combine them in a dynamic manner.} 
Specifically, given an input code snippet, we first use an IR-based technique to retrieve a similar code snippet from the corpus. 
Then we use a Cross-Encoder based classifier to decide the comment generation method to be used dynamically, i.e., if the retrieved similar code snippet is a true positive (i.e., is semantically similar to the input), 
we directly use the IR-based technique. 
Otherwise, we pass the input to the neural-based model to generate the comment.  

\textbf{Results:} We evaluate our approach on a large-scale dataset of Java projects. Experiment results show that our approach can achieve 25.45 BLEU score, which improves the state-of-the-art IR-based approach, neural-based approach, and their combination by 41\%, 26\%, and 7\%, respectively.

\textbf{Conclusions:} We propose a straightforward but effective dynamic combination of IR-based and neural-based comment generation, which outperforms state-of-the-art approaches by a substantial margin. 

\end{abstract}



\maketitle

\section{Introduction}
\label{sec:intro}

{Manually writing comments is very time-consuming, and code comments are often low-quality, missing, or mismatched after the software is upgraded~\cite{de2005study,kajko2005survey}. }
{To assist developers in writing high-quality comments or fill in absent comments,} code comment generation techniques have been proposed, which aim to generate a summary for a given code snippet automatically{~\cite{moreno2013automatic,eddy2013evaluating,iyer2016summarizing,hu2018deep,zhang2020retrieval,wei2020retrieve}}.  

Most of existing code comment generation approaches can be categorized into two orthogonal types, i.e., the information retrieval (IR) based approaches~\cite{haiduc2010supporting,haiduc2010use,edmund2014mining,wong2015clocom,kamiya2002ccfinder,li2006cp,kim2005empirical,liu2018commitmsg}, which leverage the comments of retrieved similar code snippets to generate comments for code snippets and
the neural-based approaches~\cite{iyer2016summarizing,hochreiter1997long,hu2018deep,leclair2019neural}, which treat the comment generation task as a translation problem and build neural machine translation (NMT) models to generate comments for code snippets. 
{IR-based approaches can directly leverage the existing and manually written comments, which may contain rare words or project-specific information that are difficult to be generated by NMT~\cite{koehn2017rareword}.}
In contrast, the neural-based approaches perform more robustly on general and new-coming samples with generalization capability~\cite{koehn2017rareword}.
Therefore, recent studies~\cite{zhang2020retrieval,wei2020retrieve} have gradually focused on combining the strengths of the IR-based and neural-based approaches to achieve better performance. Specifically, most of the existing approaches bind IR- and neural-based approaches statically, 
i.e., each input code sample and its retrieved similar code snippet from the IR-based approaches will be fed to the NMT model of neural-based approaches to generate comments regardless of whether the retrieved similar code snippet is actually similar to the input one or not.  
In this paper, we will refer to these approaches as \textit{IR+NMT} approaches. 
\begin{figure}[t]
    \centering
    \includegraphics[width=0.5\textwidth]{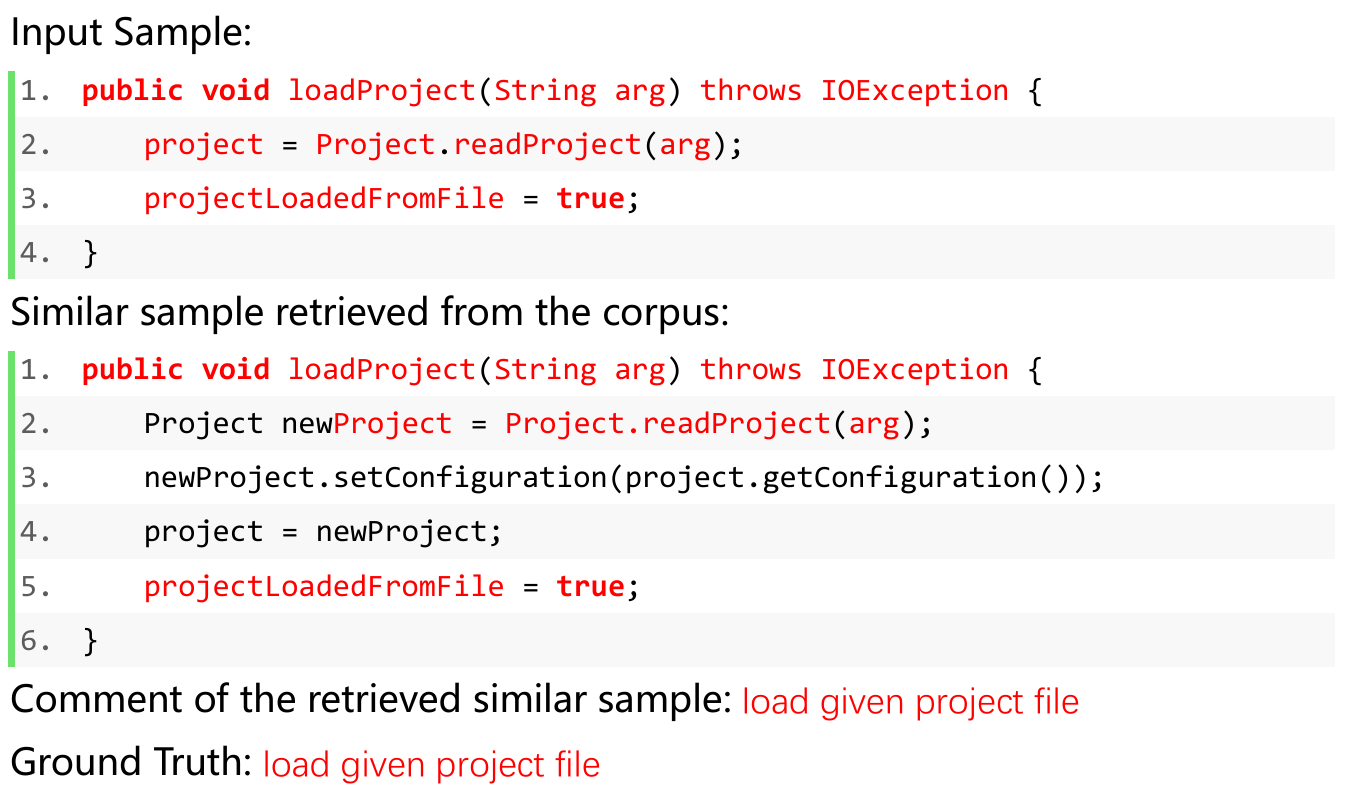}
    \caption{An example where the retrieved code snippet is semantically similar to the input one regarding both code and comment.} 
    \label{fig:IRBetter}
\end{figure}
\begin{figure}[t]
    \centering
    \includegraphics[width=0.5\textwidth]{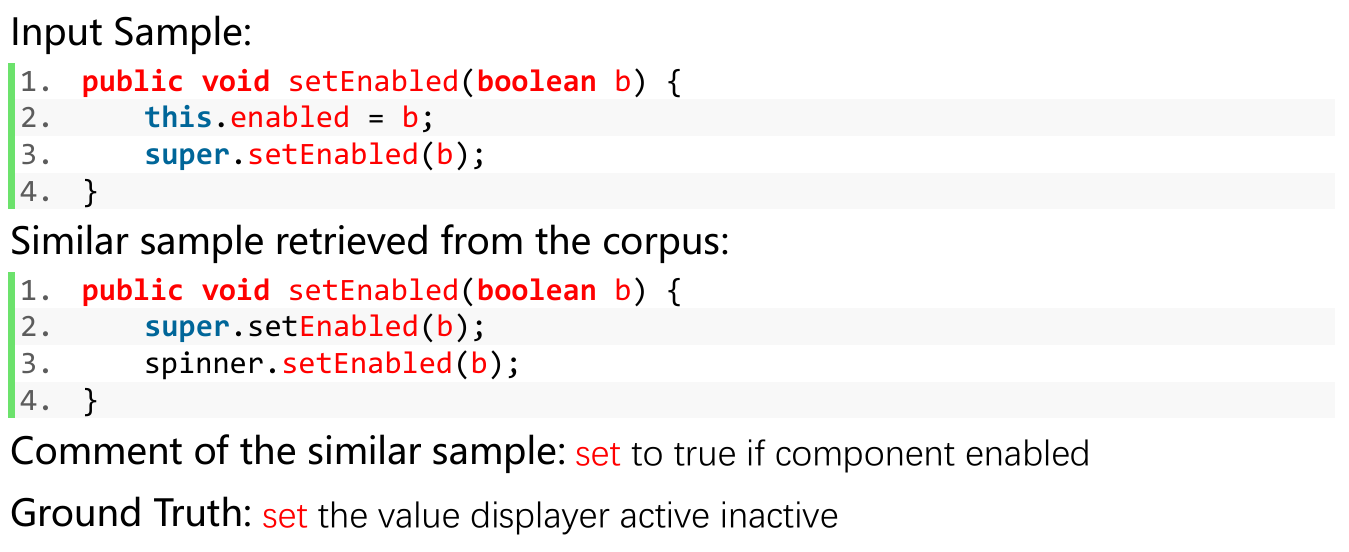}
    \caption{A false positive example where the retrieved code snippet is only textually, not semantically, similar to the input code sample.} 
    \label{fig:IRWorse}
\end{figure}

However, despite the tremendous progress of existing \textit{IR+NMT} approaches, our pilot study reveals that such a combination is not generalizable and can lead to performance degradation. 
For instances, Figure~\ref{fig:IRBetter} shows an example that the comment from the retrieved similar code snippet is a perfect match to the input code sample; thus, there is no need to feed it into the neural-based models. 
In contrast, Figure~\ref{fig:IRWorse} shows another example that a retrieved sample is highly lexical similar to the input sample in codes while they are irrelevant in comments;    
feeding the retrieved false-positive code snippets into a neural-based model will confusing the neural model and further degrade its performance. 

{In this paper, to tackle the issue of existing static binding of IR- and neural-based approaches, we propose a straightforward but effective approach to combine the strengths of the IR-based and neural-based approaches in a dynamic manner.} 
Specifically, given an input code snippet, we first use an IR-based approach to retrieve a similar code snippet from the corpus. 
Then we use a Cross-Encoder based classifier to select the comment generation method to be used dynamically, i.e., if the retrieved similar code snippet is a true positive,  
{we directly reuse the existing comment from the similar sample retrieved by the IR technique.}
Otherwise, we pass the input to the neural-based approach to generate its comment. 


To evaluate our approach, we conduct experiments on a large-scale dataset provided by LeClair et al.~\cite{leclair2019neural}, which comes from the Sourcerer repository and contains about 2M code-comment pairs. We employ BLEU~\cite{papineni2002bleu}, METEOR~\cite{banerjee2005meteor}, ROUGE-L~\cite{lin2004rouge}, and CIDER~\cite{vedantam2015cider} as evaluation metrics to evaluate predicted comments. The experimental results show that our approach can outperform state-of-the-art baselines on all selecting metrics. Specifically, our approach can achieve 25.45 BLEU score, which improves the state-of-the-art IR-based approach, neural-based approach, and their combination by 41\%, 26\%, and 7\%, respectively. 

The main contributions of this paper are as follows:
\begin{itemize}
\item We propose a straightforward but effective approach to combine the IR-based and neural-based comment generation approaches in a dynamic manner. 
\item {We have designed a Cross-Encoder based classifier, which dynamically selects the comment generation method to be used for each input sample.}


\item We conduct extensive experiments on a large-scale dataset to evaluate the performance of our approach. The experiment results show the effectiveness of our approach. 
\item We release the source code of our approach and the dataset of our experiments to help other researchers replicate and extend our study\footnote{https://zenodo.org/record/4757011}.

\end{itemize}

The rest of this paper is organized as follows. Section~\ref{sec:bg} presents the background of this study. 
Section~\ref{sec:app} describes the details of our approach. Section~\ref{sec:expsetup} and Section~\ref{sec:result} present the experiment setup and results. Section~\ref{sec:disc} discusses the strengths of our approach and threats to validity. Section~\ref{sec:rw} reviews related work. Finally, we conclude our work in Section~\ref{sec:con}.
\section{Background}
\label{sec:bg}


\subsection{Neural Machine Translation}
Recent neural-based comment generation approaches~\cite{iyer2016summarizing,hu2018deep,leclair2019neural,hu2020deep,zhang2020retrieval} treat comment generation as an end-to-end neural machine translation (NMT) task and leverage the encoder-decoder Sequence-to-Sequence (Seq2Seq) model to learn the translating pattern. Specifically, at each time step $t$, it reads one token $x_t$ from the input code snippet sequence $X=x_1,\cdots,x_n$, then the encoder updates the current hidden state $h_t$:
\begin{equation}
h_t=f(x_t,h_{t-1})
\end{equation}
where $f$ is a neural unit, e.g. GRU~\cite{cho2014gru}, LSTM~\cite{hochreiter1997long}. 

Attention mechanism~\cite{bahdanau2014attention} is adopted to focus on the critical part of the input code during decoding. For predicting target word $y_i$, a context vector $c_i$ is calculated as a weighted sum of all hidden states $h_1,\cdots,h_n$:
\begin{equation}
c_{i}=\sum_{j=1}^{n} \alpha_{i j} h_{j}
\end{equation}
The weight $\alpha_{i j}$ of each hidden state $h_{j}$ is calculated as follows:
\begin{equation}
\alpha_{i j}=\frac{\exp \left(e_{i j}\right)}{\sum_{k=1}^{n} \exp \left(e_{i k}\right)},  e_{i j}=a\left(s_{i-1}, h_{j}\right)
\end{equation}
where $s_{i-1}$ donates the last hidden state of the decoder, $a$ is an alignment model,  e.g., a Multi-Layer Perception (MLP) unit~\cite{pal1992mlp}.

At time step $i$, the hidden state $s_i$ of the decoder is updated by:
\begin{equation}
s_i=f(y_{i-1},s_{i-1})
\end{equation}
where $y_{i-1}$ is the previous generated token. Then, the decoder generates the target sequence $Y$ by sequentially predicting the conditional probability of a word $y_i$ based on the hidden state $s_i$ and the context vector $c_i$.
\begin{equation}
p\left(y_{i} \mid y_{1}, \ldots, y_{i-1}, X\right)=g\left(y_{i-1}, s_{i}, c_{i}\right)
\end{equation}
where $g$ is the generator function, e.g., a MLP layer~\cite{pal1992mlp} along with \textit{softmax}.

\begin{figure*}[t]
    \centering
    \includegraphics[width=0.94\textwidth]{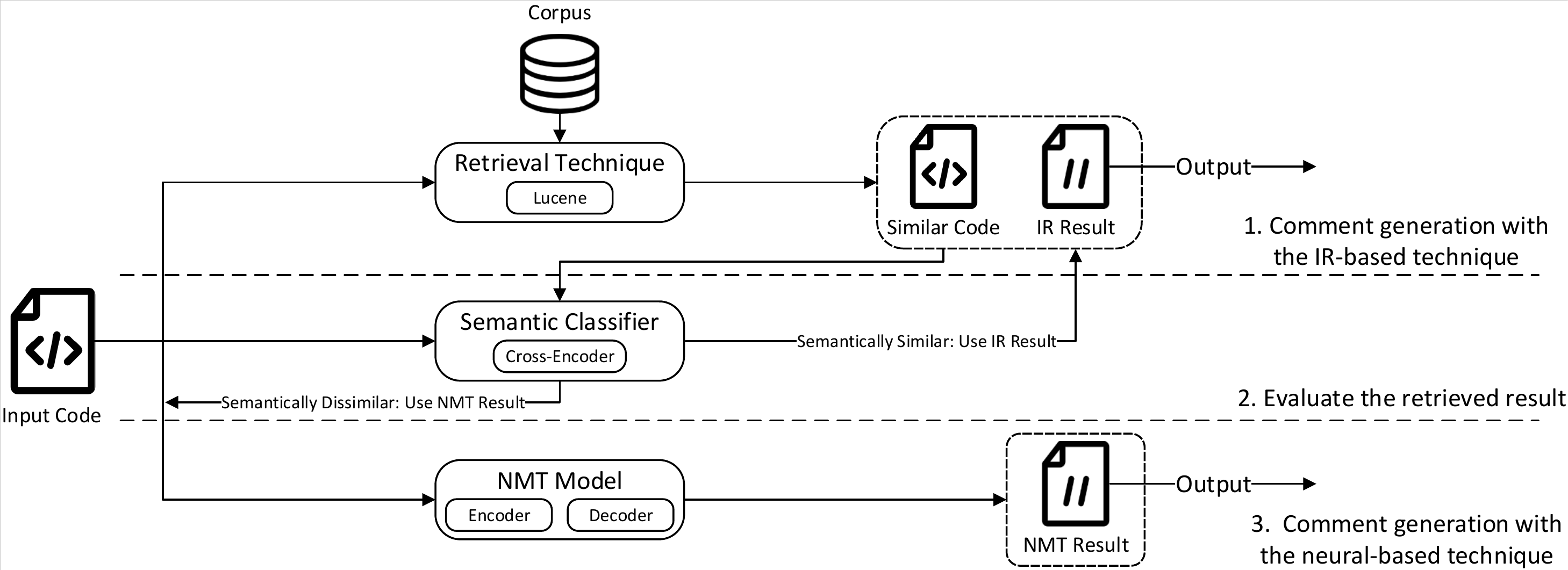}
    \caption{An overview of our approach}
    \label{fig:workflow}
\end{figure*}

The cross-entropy loss function is used to train the Seq2Seq model, i.e., minimizing the
following objective function:
\begin{equation}
\mathcal{L}(\theta)=-\sum_{i=1}^{N} \sum_{j=1}^{L} \log p\left(y_{j}^{(i)}\right)
\end{equation}
where $\theta$ donates the trainable parameters, $N$ is the number of training instances and $L$ is the length of each target sequence. $y_{j}^{(i)}$ means the $j$th word in the $i$th instance.

\subsection{Semantic Textual Similarity}
\label{sec:sts}
{To better distinguish false-positive samples, like the example shown in Figure \ref{fig:IRWorse}, we treat \textit{determining whether the retrieved results are similar to the input samples} as a supervised learning task.} The semantic textual similarity (STS) task aims to determine the semantic similarity of a given sentence pair, which is similar to our task. {The input sentence pair to the semantic classifier is the input and retrieved code snippet. The predicted label is whether the retrieved result is accurate.}


\begin{figure}[t]
    \centering
    \includegraphics[width=0.48\textwidth]{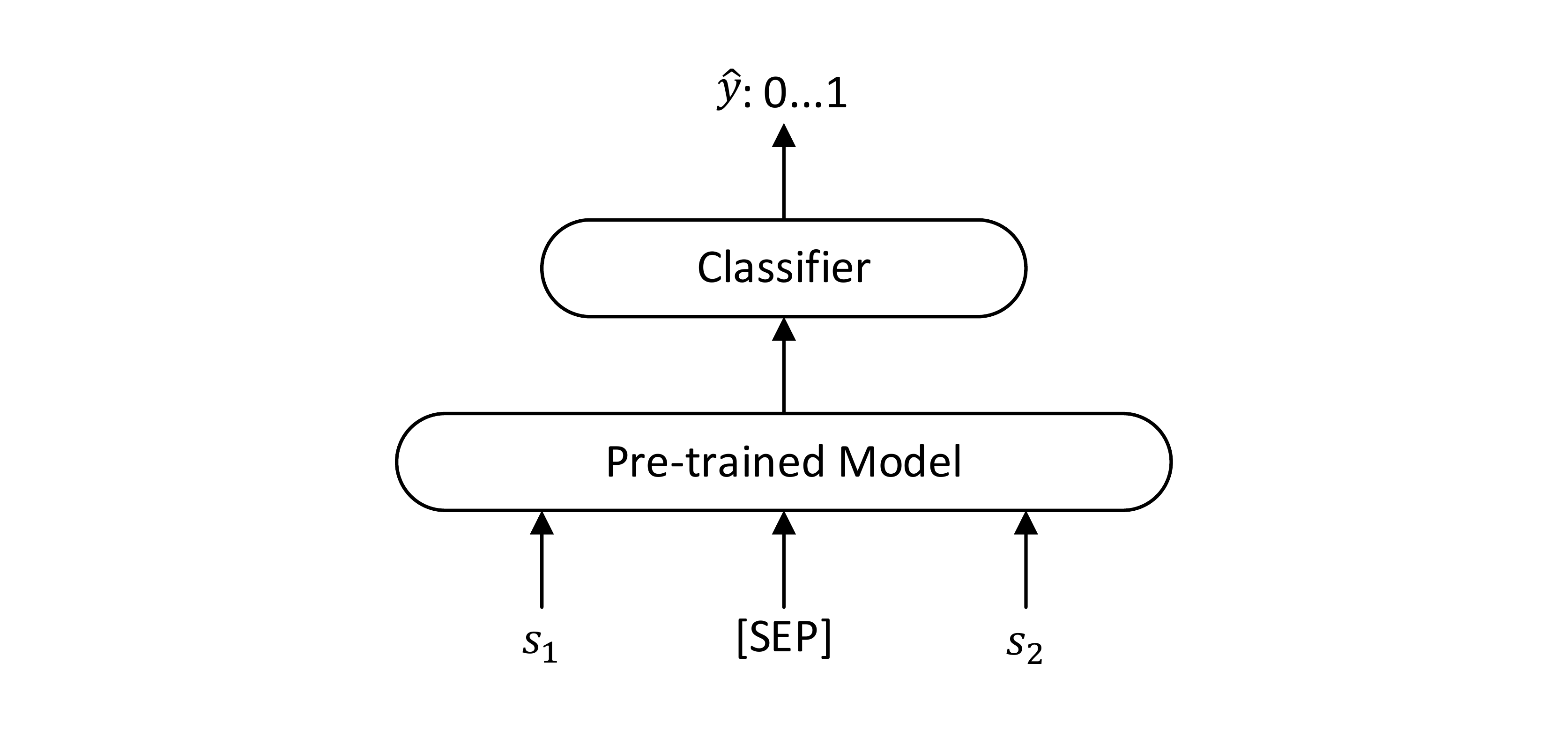}
    \caption{Structure of the Cross-Encoder}
    \label{fig:cross+siamese}
\end{figure}

Cross-Encoder~\cite{devlin2018bert} is one of the state-of-the-art methods for the semantic textual similarity (STS) task. The structure of the Cross-Encoder is shown in Figure \ref{fig:cross+siamese}.
For the given sentence pair ($s_1,s_2$),
Cross-Encoder concatenates them by a special token ([SEP]) to encode them simultaneously. A multi-head attention pre-trained model (e.g., BERT \cite{devlin2018bert}) is used to encode the concatenated sequence. In the encoding process, the self-attention mechanism allows two input sentences to perceive each other's information at a fine-grained level.
The embedding result is fed into a classifier layer that produces an output value $\hat{y}$ between 0 and 1, indicating the semantic similarity.


{In this paper, we use a Cross-Encoder based classifier to identify samples with accurate retrieved results.
For the pre-trained model of the Cross-Encoder, we choose CodeBERT~\cite{feng2020codebert}, which is trained on a large-scale code corpus consists of Java and five other programming languages~\cite{husain2019codesearchnet}. Comparing with other pre-trained models on natural language, CodeBERT can save the effort of semantic migration from natural language to programming language during fine-tuning.}

\section{Approach}
\label{sec:app}

In this work, we propose a comment generation approach that combines the strengths of the IR- and neural-based comment generation approaches dynamically. 
The key idea of our approach is straightforward:
given an input code snippet, we first use an IR-based approach to retrieve a similar code snippet from the corpus. Then we use a Cross-Encoder based classifier to select the comment generation method to be used dynamically, i.e., if the retrieved similar code snippet is a true positive, we directly use the IR result. Otherwise, we pass the input to the neural-based approach to generate the comment. 
Unlike existing IR+NMT approaches~\cite{zhang2020retrieval,wei2020retrieve}, we do not pass the information obtained by the IR-based approach to the neural network model to avoid 
textually similar but semantically dissimilar retrieved results to confuse the model. 


\subsection{Overview of Our Approach}
\label{sec:overview_app}
The workflow of our approach is shown in Figure \ref{fig:workflow}. 
Given an input sample, our approach generates its comment using the following three steps: 
1) Comment generation with the IR-based technique (Section \ref{approach:stage1}). In this step, our approach extracts the comment from the most similar sample retrieved from the corpus through the {IR-based} retrieval technique. 
2) Evaluate the retrieved result 
(Section \ref{approach:stage2}).  We use a Cross-Encoder based classifier to determine whether the retrieved code snippet is similar to the input semantically. We assume that directly leveraging the existing comment from a true-positive similar sample{, which may contain low-frequency words and project-specific information that hard to be generated by NMT~\cite{koehn2017rareword,zhang2020retrieval,wei2020retrieve},} will be more accurate and informative than the generated result of NMT models. Therefore, when the retrieved code snippet is similar to the input, our approach will reuse the comment of the retrieved code snippet. Otherwise, we assume that the current sample needs to be inferred by generation-based methods.
3) Comment generation with the neural-based technique (Section \ref{approach:stage3}). 
For the input sample whose retrieval result is determined to be inaccurate from the previous step, the neural model is used to automatically generate its  comment based on the input code snippet and corresponding AST sequence. 


\subsection{Comment Generation with The IR-based Technique}
\label{approach:stage1}
{This step aims to provide an existing comment for each input sample that may be reusable from the retrieved similar code snippet.}

To identify the most similar sample for a given sample,  
in this work, we reuse the retrieval method of \ReCom ~\cite{wei2020retrieve}, which is a code lexical similarity based retrieval method. The retrieval module of \ReCom uses the training set as the corpus. It retrieves the sample with the highest lexical similarity between code snippets based on BM25 algorithm from search engine Lucene\footnote{https://lucene.apache.org/}, a widely used similarity metric. For each term in the given code snippet, its relevance score to the candidate code snippet is calculated based on the term frequency. Then, the BM25 score between the input and candidate code snippet is calculated as a weighted sum of the relevance score of each term, where the weight of each term is calculated based on its inverse document frequency. Finally, the candidate code snippet with the highest BM25 score is selected as the retrieved result. Note that, IR-based approach does not have a training process. We use the settings of BM25 from \ReCom to run our experiments. 



\subsection{Evaluate The Retrieved Result with The Cross-Encoder based Classifier}
\label{approach:stage2}
In the previous step, we have provided an existing comment {from the retrieved similar code snippet} for each input sample. 
However, as shown in Figure ~\ref{fig:IRWorse}, the results of the IR technique could be incorrect, thus to 
achieve more accurate determination, we compare the semantic between the input and the retrieved code snippet by a semantic model to predict whether the IR result is accurate and can be directly reused.
To identify samples with accurate IR results, we compare the input with the retrieved code snippet semantically rather than textually. This is because determining the performance of IR results from text similarity is not accurate enough. As shown in Figure~\ref{fig:IRWorse}, the input and the retrieved code snippet are very similar, with only 2-3 tokens different. However, their corresponding comments have only one token in common. In this work, we use the Cross-Encoder model for the semantic comparison, one of the state-of-the-art methods for the semantic textual similarity (STS) task. Figure  \ref{fig:cross+siamese} shows the structure of the Cross-Encoder. 
The input to the model is the input and retrieved code snippet. Two snippets are concatenated into a sequence through a specific token [SEP] provided by BERT~\cite{devlin2018bert} and simultaneously passed to a pre-trained multi-level transformer ~\cite{vaswani2017attention} network for embedding. We choose CodeBERT~\cite{feng2020codebert} as the pre-trained model to save the effort of semantic migration. The embedding result of the two snippets is fed into a liner classifier layer that produces an output value between 0 and 1, indicating the degree of semantic similarity:
\begin{equation}
\hat{y}=T(code_{input}, code_{retrieved})W
	\label{eq_cross_pred}
\end{equation}
where $\hat{y}$ is the predicted degree of semantic similarity, $W$ is the weight of the linear layer, and $T(code_{input}, code_{retrieved})$ is the embedding result of the input and retrieved code snippet.

The training process is fine-tuning the semantic model with pairs of code snippets to the target that if a semantically similar snippet is retrieved, the model returns 1, otherwise returns 0. We use the classic cross-entropy loss function to fine-tune the model:
\begin{equation}
Loss=-(y*log(\hat{y})+(1-y)*log(1-\hat{y}))
    \label{eq_cross_loss}
\end{equation}
where $y$ indicates the golden label of whether the retrieved result is accurate. 

The details of how we train the Cross-Encoder based classifier are available in Section~\ref{setup:cross}.

\subsection{Comment Generation with The Neural-based Technique}
\label{approach:stage3}
In the previous step, we have identified samples with accurate IR results. While the remaining input samples, we further use the neural-based approach to generate comments for them.
Specifically, in this step, we first build and train an NMT model on our corpus. 
Then we input samples that are determined to have inaccurate IR results in the previous step to this model to generate comments. This step aims to use the generalization ability of NMT to generate comments for general input samples.

{In this step, we use the state-of-the-art neural-based comment generation method, i.e., } DeepCom~\cite{hu2020deep}.
DeepCom is an encoder-decoder structure model with the attention mechanism~\cite{bahdanau2014attention}. The input of the model contains both code and AST sequences, where the code sequence contains semantic information such as identifier names, and the AST sequence contains structural information. Using semantic and structural information from the input code snippet simultaneously can help the model understand them more clearly and predict more accurately~\cite{hu2020deep}. The model uses two encoders to encode the code sequence and the AST sequence, respectively.
We follow the model training and turning processes described in DeepCom~\cite{hu2020deep} to re-train the models on our corpus (details are in Section~\ref{sec:trainDeepCom}).
\section{Experiment Design}
\label{sec:expsetup}
\subsection{Dataset}
We use the FunCom dataset provided by LeClair et al.~\cite{leclair2019neural} to conduct our experiments, which has been used in many existing studies~\cite{leclair2019neural,wei2020retrieve,haque2021action}. The FunCom dataset is collected from a large Sourcerer repository ~\cite{lopes2010uci}, which contains over 50,000 projects and 5.1 million java methods. LeClair et al. treat the first sentence of the Javadoc of each method as its comment~\cite{kramer1999javadoc}, use srcML~\cite{collard2011srcml} to extract AST sequences from source codes, then serialize them by the SBT method proposed by Hu et al.~\cite{hu2018deep}. To reduce the vocabulary size, LeClair et al. adopt a series of preprocessing to the code and comment text: splitting identifiers in code and comment by camel case and underscore, removing non-alpha characters (including symbols) from the text, and converting the text to lowercase. To better simulate the case where only AST is known, identifiers in the AST sequence are replaced with <OTHER>. To reduce duplicate samples between the training and test set, LeClair et al. use a heuristic rule~\cite{shimonaka2016removeauto} to filter out auto-generated codes which are very similar to each other and too easy to be learned and predicted by the model. In addition, LeClair et al. divide all the data by project in the dataset building stage: data from 90\% of projects are divided as the training set, 5\% as the validation set, and 5\% as the test set. After filtering, the FunCom dataset has about 2M code-comment pairs for training and testing.

{The FunCom dataset is the most reasonable dataset to the best of our knowledge, which has a large amount of data and excludes noisy data, thus allowing us to evaluate the model's generalization ability more accurately.}

\subsection{Experiment Settings}
In this work, we train 
both DeepCom~\cite{hu2020deep} 
and the Cross-Encoder based  classifier~\cite{devlin2018bert} on the FunCom dataset. Their training details are as follows.

\subsubsection{Training Details of DeepCom}
\label{sec:trainDeepCom}
We use the default settings of DeepCom for training, i.e., the encoder and decoder use a single-layer Gated Recurrent Unit (GRU) structure~\cite{cho2014gru}. Both the word embeddings and the GRU hidden states are set to 256. In the decoding stage, beam search~\cite{wiseman2016beamsearch} is leveraged to obtain more accurate results, with the beam width is set to 5. 
We use the entire FunCom dataset for training and validation. DeepCom is trained on the FunCom training set (19,548,008 samples in total). {Following DeepCom, we use Stochastic Gradient Descent (SGD) based optimizer to train the model,} 
the initial learning rate is set to 0.5, and the learning rate decay factor is set to 0.95. In addition, to save GPU memory, we set the batch size to 256. Every 2000 training steps, the checkpoint is saved and validated on the FunCom validation set (104,273 samples in total). After 20 epochs of training (about 150,000 steps), the best parameters are selected from the checkpoint that performs best on the validation set. We trained the model on a Linux server with the NVIDIA RTX 2060S GPU with 8GB memory, which took about 70 hours for training.   

\subsubsection{Training Details of Cross-Encoder Based Classifier}
\label{setup:cross} 
We use the Sentence-Bert~\cite{reimers2019sentencebert} package to build and train the Cross-Encoder based classifier. In order to save the effort of language semantics migration, we adopt the widely used CodeBERT pre-trained model~\cite{feng2020codebert}, a 24-layer bidirectional  transformer~\cite{vaswani2017attention} network. 

To label the dataset for training the Cross-Encoder based classifier, we use code-comment pairs from the validation set of FunCom (104,273 samples in total). For each sample, we use the IR-based approach (details are in Section~\ref{approach:stage1}) to retrieve the most similar code snippet, and the corresponding comment will be treated as the IR result. Then we use a trained neural model (i.e., DeepCom) to generate its comment, i.e., NMT result. 
The label of the sample is \textit{whether the IR result is more accurate}. 
Specifically, we use  \textit{sentence\_bleu} metric in the NLTK~\cite{loper2002nltk} package to calculate the similarities of the IR result and NMT result with ground truth, respectively. If the score of the IR result is greater than the score of the NMT result, it is labeled as a positive sample; otherwise, it is labeled as a negative sample. We further exclude cases where both methods perform poorly from positive samples (e.g., both IR result and NMT result fail to hit any word in the ground truth comment). Finally, we obtain a triplet for each sample: \textit{< Input code snippet, Retrieved code snippet, Is\_IR\_Result\_Better? >}. After labeling  the data, we take 90\% of triplets (93,846 samples) for training, and the remaining 10\% (10,427 samples) of triplets are used as a developmentset for tuning the parameters and testing. 

We use Adam optimizer~\cite{kingma2014adam} to train the Cross-Encoder based classifier, and the initial training rate is set to 2e-5, the learning rate decay factor is set to 0.99. We set the batch size to 16, and for every 2000 training steps, save the checkpoint and validate it on the development set. After fine-tuning 5 epochs (about 55,000 steps), the best parameters are selected from the checkpoint that performs best on the development  set. We fine-tuned the model on a Linux server with the NVIDIA Titan RTX GPU with 24GB memory, which took about 3 hours for fine-tuning.

\subsection{Baselines}
\label{setup:combaselines}

\subsubsection{Baselines for Evaluating Our Comment Generation Approach}
\label{setup:baselines}

To investigate the performance of our comment generation method, we selected the IR-based approach (details are in  Section~\ref{approach:stage1}), four state-of-the-art neural-based comment generation methods~\cite{zhang2020retrieval,leclair2019neural,hu2020deep}, and two state-of-the-art IR+NMT methods~\cite{zhang2020retrieval,wei2020retrieve} as baselines.


\vspace{0.05cm}
\noindent \textbf{1) Neural-based methods}
\vspace{0.05cm}


\textbf{Rencos NMT module}~\cite{zhang2020retrieval} is the NMT module of Rencos~\cite{zhang2020retrieval}, a standard attentional Seq2Seq model where the encoder is bidirectional LSTM and the decoder is LSTM. This baseline represents a fundamental solution to use NMT on code to comment problem, i.e., train an NMT with code as input and comment as output. 

  \textbf{attendgru}~\cite{leclair2019neural}
 is an attentional Seq2Seq-like model. This baseline predicts only one word at a time. In the encoding process, the model encodes both the code sequence and the output sequence predicted in previous steps. In the decoding process, the model predicts the next most likely word and appends it to the output sequence for the subsequent prediction steps. 

 \textbf{ast-attendgru}~\cite{leclair2019neural} 
 is also an attentional Seq2Seq-like model. This baseline adds AST as an additional input to improve the prediction performance. LeClair et al.~\cite{leclair2019neural} use the traversal method SBT~\cite{hu2018deep} to flatten the AST into a sequence and adds an additional encoder for the AST sequence.

\textbf{DeepCom}~\cite{hu2020deep} is a standard attentional Seq2Seq model, where the encoder and the decoder are both Gated Recurrent Unit (GRU). The inputs to the model are code and AST sequences. As our proposed method takes the prediction results of this baseline as the NMT results, improvement from combining IR results can be directly measured by comparing the performance of our proposed method with this baseline.

\vspace{0.05cm}
\noindent \textbf{2) IR+NMT methods}
\vspace{0.05cm}


\textbf{Rencos}~\cite{zhang2020retrieval} combines the IR-based and neural-based comment generation by feeding the most semantic-level and syntactic-level similar code snippets of an input code snippet retrieved by IR-based approach into the neural-based approach to generate the comment. Specifically, given an input code snippet, Rencos retrieves its two most similar code snippets on semantic-level and syntactic-level. Then, the input code snippet and its two similar ones are fed separately into a trained code-to-comment NMT model to generate the comment.

\textbf{\ReCom}~\cite{wei2020retrieve} uses additional encoders to encode information from the retrieved sample of IR-based approaches. For a given code snippet, a similar sample with the highest text similarity is retrieved from the corpus. Then \ReCom takes the given code, its AST, code, and comment of the similar sample as input and encodes them by four different encoders. The encoding results are fused by the similarity between the input and the retrieved code and then passed to the decoder to obtain the predicted comment. 
 
\subsubsection{Baselines for Evaluating Cross-Encoder Based Classifier}
\label{setup:clasbaselines}

To evaluate the effectiveness of our Cross-Encoder based classifier (details are in Section~\ref{approach:stage2}) in determining whether IR results are accurate, we adopt two other classification methods as the baselines. 



\textbf{Lexical-level Similarity} is a simple method determining whether the IR result is accurate based on the lexical similarity between the input and retrieved code. If the similarity is greater than an appropriate threshold, we assume that the IR result is accurate and treat it directly as the output; otherwise, the neural-based approach will be used to generate its comment. We follow~\cite{gros2020code} and use the \textit{sentence\_bleu} metric in the NLTK~\cite{loper2002nltk} package to calculate the lexical similarity. This method does not require training but needs to determine an appropriate threshold that makes the dynamic combination of IR- and neural-based approaches on the test dataset can achieve optimal performance. 
To find the optimal threshold, we experiment the threshold values from 0 to 1 with an interval of 0.05. 
When the threshold value is 0.40, this approach achieves optimal performance on FunCom's validation set. 
Thus, we use 0.4 as the threshold value in our experiments.


\textbf{Siamese Network}~\cite{bromley1993signature} is another state-of-the-art method on the semantic textual similarity (STS) task.
It consists of two identical encoders to encode the two input sentences separately, which share the same model structure and parameters. Then, the distance between two embeddings is treated as the semantic similarity between the sentence pair.
We use the implementation from GitHub\footnote{https://github.com/tlatkowski/multihead-siamese-nets} to build a Siamese network, which uses a bidirectional LSTM (Bi-LSTM)~\cite{schuster1997bilstm} with 256 hidden sizes as the encoder structure and chooses manhattan distance as the similarity of embedding vector of input sentence pairs. Like Cross-Encoder, we use the labeled dataset described in Section \ref{setup:cross} to train the Siamese network.
\begin{table*}[t]
\caption{The performance of our method compared with other comment generation baselines (the best ones are marked in bold). 
The percentages in parentheses indicate the relative improvement achieved by our method compared to the IR-based method and NMT-based method (DeepCom), respectively. 
}
\centering
\resizebox{\textwidth}{!}{
\begin{tabular}{@{}clllllllll@{}}
\toprule
\textbf{Type}                           & \multicolumn{1}{c}{\textbf{Approach}} & \textbf{BLEU(\%) }                     & \textbf{BLEU\textsubscript{1}(\%) }                    & \textbf{BLEU\textsubscript{2}(\%)}                      & \textbf{BLEU\textsubscript{3}(\%)}                     & \textbf{BLEU\textsubscript{4}(\%)}                     & \textbf{METEOR(\%)}                     & \textbf{ROUGE-L(\%)}                     & \textbf{CIDER}                     \\ \midrule
IR-Based                        & Re2Com Retrieve   Module        & 18.04                      & 32.04                     & 17.84                      & 14.4                      & 12.88                     & 15.41                      & 30.64                     & 1.643                     \\
\midrule  \hline
\multirow{4}{*}{Neural-based}      &  Rencos NMT Module                  & 19.15                      & 34.64                     & 20.58                       & 15.11                     & 12.49                     & 18.92                      & 39.54                     & 2.074                     \\
                               & attendgru                       & 19.26                      & 38.64                     & 21.71                      & 14.63                     & 11.21                     & 19.34                      & 40.16                     & 1.984                     \\
                               & ast-attendgru                   & 19.73                      & 39.8                      & 22.25                      & 14.93                     & 11.46                     & 19.43                      & 39.94                     & 1.952                     \\
                               & DeepCom                         & 20.11                      & 40.71                     & 22.57                      & 15.17                     & 11.73                     & 19.92                      & 40.25                     & 2.044                     \\
\midrule \hline
\multirow{2}{*}{IR+NMT}        & Rencos                          & 19.86                      & 36.7                      & 21.58                      & 15.55                     & 12.64                     & 19.17                      & 39.9                      & 2.066                     \\
                               & \ReCom                          & 23.69                      & 40.38                     & 24.74                      & 19.12                     & 16.48                     & 20.28                      & 39.91                     & 2.282                     \\
\midrule  \hline
\multicolumn{1}{l}{Our Method} &  & \textbf{25.45 (41\%/26\%)} & \textbf{43.92 (37\%/7\%)} & \textbf{27.08 (51\%/19\%)} & \textbf{20.38(41\%/34\%)} & \textbf{17.3 (34\%/47\%)} & \textbf{22.03 (42\%/10\%)} & \textbf{43.21 (41\%/7\%)} & \textbf{2.46 (49\%/20\%)} \\ \bottomrule
\end{tabular}
}
\label{tab:rq1_bleu}
\end{table*} 

\subsection{Evaluation Metrics}
\label{sec:emetrics}
\subsubsection{Metrics for Evaluating Generated Comments}
\label{setup:commetrics}

In our experiments, we follow Rencos~\cite{zhang2020retrieval} and evaluate the performance of different comment generation methods with four common metrics, i.e., BLEU~\cite{papineni2002bleu}, METEOR~\cite{banerjee2005meteor}, ROUGE-L~\cite{lin2004rouge}, and CIDER~\cite{vedantam2015cider}, which are widely used in machine translation~\cite{sutskever2014machine}, text summarization~\cite{rush2015summarization}, and image captioning{~\cite{you2016image}}.

\textbf{BLEU}~\cite{papineni2002bleu} measures the similarity between the generated comment and ground truth by the geometric mean of $n$-gram matching precision scores $p_n$. A brevity penalty $BP$ is used to prevent very short generated sentences.
\begin{equation}
BLEU=BP \cdot \exp \left(\sum_{n=1}^{N} w_{n} \log p_{n}\right)
\end{equation}
where $w_{n}$ is the uniform weight, and $N$ is set to 4 in our paper. We report a composite BLEU score in addition to BLEU\textsubscript{1} through BLEU\textsubscript{4} in our experiment.

\textbf{METEOR}~\cite{banerjee2005meteor} calculates the similarity scores by the unigram precision $P$ and recall $R$, and multiplied by a penalty of language order:
\begin{equation}
METEOR=\left(1-\gamma \cdot frag^{\beta}\right) \cdot \frac{P \cdot R}{\alpha \cdot P+(1-\alpha) \cdot R}
\end{equation}
where $frag$ is the fragmentation fraction. $\alpha$, $\beta$, and $\gamma$ are three parameters whose default values are 0.9, 3.0 and 0.5, respectively.

\textbf{ROUGE-L}~\cite{lin2004rouge} is calculated by the Longest Common Subsequence (LCS) matching F-score. Suppose the length of the target sentence ($X$) and the predicted sentence ($Y$) are m and n, respectively, and the length of the LCS between them is $LCS(X, Y)$, then:
\begin{equation}
\small
P_{lcs}=\frac{LCS(X, Y)}{n}, R_{lcs}=\frac{LCS(X, Y)}{m}, F_{lcs}=\frac{\left(1+\beta^{2}\right) P_{lcs} R_{lcs}}{R_{l c s}+\beta^{2} P_{l c s}}
\end{equation}
where $F_{lcs}$ is the value of ROUGE-L, $P_{lcs}$ and $R_{lcs}$ denote the LCS precision and recall, respectively, and $\beta=P_{lcs}/R_{lcs}$.

\textbf{CIDER}~\cite{vedantam2015cider} examines whether the prediction result has captured the critical information. Given the generated summary $c_{i}$ and the ground-truth $s_{i}$, CIDER is calculated by the frequency of $n$-grams and TF-IDF weighting:
\begin{equation}
\begin{aligned}
CIDER_{n}\left(c_{i}, s_{i}\right)=\frac{1}{M}*\sum_{j=1}^{M}\frac{g^{n}\left(c_{i}\right)* g^{n}\left(s_{i j}\right)}{\left\|g^{n}\left(c_{i}\right)\right\| *\left\|g^{n}\left(s_{i j}\right)\right\|} \\
CIDER\left(c_{i}, s_{i}\right)=\sum_{n=1}^{N} w_{n} CIDER_{n}\left(c_{i}, s_{i}\right)
\end{aligned}
\end{equation}
where $N$ is set to 4, $g^{n}\left(c_{i}\right)$ denotes the TF-IDF weight vector of all $n$-gram in sentence $c_{i}$, $M$ represents the number of reference sentences for each sample (in our work, $M = 1$) . The final result $CIDER\left(c_{i}, s_{i}\right)$ is calculated by summing of the scores for different $n$-grams ($CIDER_{n}\left(c_{i}, s_{i}\right)$) with weight $w_n$.


\subsubsection{Metrics for Evaluating Cross-Encoder Based Classifier}
\label{setup:clasmetric}
To evaluate whether the classifier can accurately distinguish samples with accurate IR results, we use four metrics commonly used in classification problems to verify the performance of our Cross-Encoder based classifier and baselines, i.e., accuracy, precision, recall, and F1-score. 
\begin{equation}
\begin{aligned}
Accuracy=\frac{TP+FN}{TP+FP+TN+FN} \\ 
Precision=\frac{TP}{TP+FP},Recall=\frac{TP}{TP+FN} \\
F1\_score=\frac{ Precision \cdot Recall }{Precision+Recall}
\end{aligned}
\end{equation}
where $TP$/$FP$ donates the number of \textit{positive samples identified by the classifier that are/are not samples with accurate IR results}, and $TN$/$FN$ donates the number of \textit{negative samples identified by the classifier that are/are not samples with inaccurate IR results}.

\subsection{Research Questions}
We perform a large-scale comparative study to answer the following three research questions for evaluating our approach. 

\begin{itemize}

\item \textbf{RQ 1 (Performance):} How does our approach compare to the commonly-used and state-of-the-art comment generation baselines?

\vspace{4pt} 

\item \textbf{RQ 2 (Accuracy of classification):} What is the accuracy of our Cross-Encoder based classifier? 
\vspace{4pt}

\item \textbf{RQ 3 (Generalizability)}: Does our approach work with other NMT methods?
\end{itemize}

In RQ1, we set out to investigate the performance of generated comments of our proposed approach by comparing with seven state-of-the-art baselines (details are in Section \ref{setup:baselines}). In RQ2, we evaluate whether the Cross-Encoder based classifier can effectively distinguish samples with accurate retrieved results by comparing with two baselines (details are in Section \ref{setup:clasbaselines}). In RQ3, we explore whether our approach is applicable for other neural comment generation approaches, i.e., still can obtain a significant improvement from dynamically combining with IR results.
\section{Result Analysis}
\label{sec:result}
\subsection{RQ 1: Our Approach vs.  Baselines}
\label{sec:rq1}

{\textbf{Experimental Method.} To answer this research question, we compare our approach with comment generation baselines listed in Section \ref{setup:combaselines}. 
All baselines are trained on the FunCom  training set.
We compare generated comments of our approach and other baselines on the FunCom test set by four evaluation metrics described in Section \ref{setup:commetrics}.} 

\textbf{Result.} Table~\ref{tab:rq1_bleu} shows the performance of our method compared to other comment generation baselines.
{Overall, } 
our approach achieves the best performance on all evaluation metrics. 
Our approach achieves a 26\% improvement on BLEU and a 7\%-47\% improvement on other metrics compared to DeepCom, the state-of-the-art neural-based approach{, and achieves a 7\% improvement on BLEU compared to Re\textsuperscript{2}Com, the state-of-the-art IR+NMT approach.}

From the table, we can see that the IR-based approach has a similar performance to neural-based approaches.
The IR-based approach achieves 18.04 BLEU score, while neural-based approaches perform slightly better than it and achieve BLEU score range from 19.15 to 20.11. One of the possible reasons that the neural-based approaches and the IR-based approach perform similarly can be that the word distributions in the training and test datasets are different. Some custom identifiers in the test set samples may be rare or even absent from the training set, making it hard for the model to capture their information accurately~\cite{karampatsis2020big}.

For the two existing combinations of IR-based and neural-based approaches, i.e., Rencos and Re\textsuperscript{2}Com, as we can see from the table, both could outperform IR-based and neural-based approaches. 
Specifically, Rencos achieves 19.86 BLEU score by fusing prediction results of the input code snippets with similar snippets. Re\textsuperscript{2}Com achieves 23.69 BLEU scores by feeding the codes and comments of similar samples into the neural model. Our method achieves a higher 25.45 BLEU score by dynamically combining IR results and NMT results. In addition, both Rencos and \ReCom fail to improve the performance of the METEOR and ROUGE-L metrics significantly, but our approach achieves a significant improvement.


{We have also conducted the Wilcoxon signed-rank test~\cite{wilcoxon1963wilcoxon} $(p<0.05)$ to compare the performance of our approach and these baselines. The test result suggests that our approach achieves significantly better performance than baseline approaches in BLEU, METEOR, ROUGE-L, and CIDER.}

\mybox{Our approach significantly outperforms the state-of-the-art comment generation baselines. The improvements on the IR-based approach, neural-based approach, and their combination are 41\%, 26\%, and 7\% in terms of BLEU score, respectively.}
\subsection{RQ 2: Cross-Encoder  vs. Other Classification Algorithms}
\label{sec:rq2}

\textbf{Experimental Method.} To answer this research question, we compare the Cross-Encoder based classifier with other classifier baselines listed in Section \ref{setup:clasbaselines}. 
{Specifically, we apply these approaches on the test set labeled as described in Section~\ref{setup:cross} and use accuracy, precision, recall, and F1-score to measure the performance. In addition, we replace the Cross-Encoder based classifier of our approach with other classifier baselines, then use BLEU to measure the quality of the generated comments.} 


\textbf{Result. }
The performance of each classification method is shown in Table \ref{tab:rq2_class}. 
{Overall, our approach (the Cross-Encoder based classifier) outperforms the two baselines on all the five metrics.}


\begin{table}[t]
\caption{The performance of different classification algorithms}
\centering

\resizebox{\linewidth}{!}{
\begin{tabular}{@{}llllll@{}}
\toprule
\multirow{2}{*}{Approach}               & \multicolumn{4}{c}{Classification Performance}  &  Generated Comments \\ \cmidrule(l){2-6}
                                      & Accuracy(\%)       & Precision(\%)      & Recall(\%)         & F1-score(\%)      & BLEU(\%)                             \\ \midrule
lexical-level similarity              & 71.3          & 65.1          & 37.8          & 47.9          & 24.22                            \\
Siamese Network & 68.9          & 59.2          & 34.4         & 43.5          & 23.5                            \\
\midrule  \hline
Cross-Encoder                           & \textbf{73.6} & \textbf{70.2} & \textbf{41.9} & \textbf{52.5} & \textbf{25.45}                   \\ \bottomrule
\end{tabular}
}
\label{tab:rq2_class}
\end{table}
\label{rq:rq2_text}
The first row of Table \ref{tab:rq2_class} shows the performance of the lexical-level similarity method (details are in Section \ref{setup:clasbaselines}), which achieves an accuracy of 71.3\% in inferring whether 
{the IR results are accurate}. Its combined results achieve 24.22 BLEU score, which is better than Re\textsuperscript{2}Com. 
Significant improvement can also be achieved even without training a classifier for comparison, which further validates that our idea of dynamically combining IR results with NMT results is indeed practical. However, the text-similarity-based approach also 
{suffers the issues of false-positive as} shown in Figure \ref{fig:IRWorse}. To identify such false-positive samples, 
we use the Cross-Encoder, a semantic-based classifier, to more accurately predict whether the IR results are{ accurate}. 

The second row of Table \ref{tab:rq2_class} shows the performance of the Siamese network method (details are in Section \ref{setup:clasbaselines}). We train a Bi-LSTM network with strong expressive capability from scratch to determine semantics similarity. However, the Siamese network does not perform as well as expected; its performance is even worse than the lexical-level similarity method we showed above. 
%
{One possible reason is that the model focuses on irrelevant features instead of the semantic gap between code snippet pair, leading to over-fitting and poor performance.} 

The third row of Table \ref{tab:rq2_class} shows the performance of our Cross-Encoder based classifier. 
{Overall, our Cross-Encoder based classifier achieves the best performance on all metrics. The high accuracy (73.6\%) and precision (70.2\%) 
validate that it can help achieve our goal of filtering false-positive retrieval results, i.e., textually similar but semantically dissimilar. Besides, we can also see that the performance of the combined result increases with the increase of accuracy of the classification, which suggests that the performance of our comment generation approach can be improved by}
{better distinguishing samples with accurate IR result.}


\mybox{Our Cross-Encoder based classifier can accurately identify samples with accurate IR results. 
Besides, our idea of {dynamically combining} IR-based and neural-based approaches can outperform the state-of-the-art IR+NMT approaches even with the naive textual-similarity algorithm.}

\subsection{RQ 3: Generalizability}
\label{sec:rq3}

\textbf{Experimental Method.} 
Different neural models might generate different results, which can affect the generalizability of our approach. 
To evaluate the generalizability of our approach, we replace the DeepCom in our approach with three other neural-based baseline approaches (listed in Section~\ref{setup:combaselines}). 
Then we measure the quality of generated comments with BLEU.
\begin{table}[t]
\caption{The performance (BLEU) of different NMT results combined with IR results.  The percentages in parentheses indicate the relative improvement achieved by combining with IR results}
\centering
\footnotesize
\begin{tabular}{@{}llll@{}}
\toprule
Approach    & NMT Only  & Combined Result & Improvement \\ \midrule
Rencos NMT Module & 19.15    & 24.95   & 5.8 (30\%)  \\
attendgru      & 19.26    & 25.32   & 6.06 (31\%) \\
ast-attendgru  & 19.73    & 25.34   & 5.61 (28\%) \\ \hline \hline
DeepCom       & 20.11    & 25.45   & 5.34 (26\%) \\ \bottomrule
\end{tabular}
\label{tab:rq3_bleu}
\end{table}

\textbf{Result.} Table \ref{tab:rq3_bleu} shows the performance of other neural-based approaches combined with IR results.  
Overall, after combining IR results, all three neural methods achieve better performance with 24.95-25.34 BLEU score. 
Specifically, Rencos NMT module, attendgru, and ast-attendgru can achieve relative improvements of 30\%, 31\%, and 28\% from combining IR results, respectively, which are even higher than the relative improvement of DeepCom (26\%). The above results fully demonstrate that the performance of our proposed approach remains stable across different neural approaches.  Moreover, all the combined results outperform \ReCom, the current state-of-the-art IR+NMT method, which again validates the feasibility of our idea of dynamically combining IR results and NMT results.

\mybox{The performance of our approach remains stable across different neural-based comment generation approaches.} 
\section{Discussion}
\label{sec:disc}
\subsection{Why Our Approach Performs Better?}
To investigate why our proposed approach can achieve better performance, we partition the 90,908 samples in the test set into two sets, i.e., 
samples on which the IR-based approach performs better (IR-better samples) and samples on which the neural-based approach (DeepCom) performs better (NMT-better samples). 
Overall, there are 31,636 samples (34.8\%) where the IR-based approach performs better, and 59,272 samples (65.2\%) where the neural-based approach performs better. We then recalculate the performance (based on BLEU) of the four methods in these two sets, i.e., \ReCom retrieve module (IR-based approach), DeepCom (neural-based approach), ReCom (IR+NMT approach), and our approach. The results are in Table~\ref{tab:diss_2part}.

From the table, we can see that
for IR-better samples, 
the IR-based approach, i.e., \ReCom retrieve module, can directly leverage existing comments from similar samples in the corpus and achieves 39.55 BLEU score, which is almost twice as large as the score of the neural-based approach,  i.e., DeepCom. For NMT-better samples, since no similar sample can be retrieved from the corpus, the IR-based approach performs poorly on these general samples and only achieves 5.25 BLEU score. In contrast, the neural-based approach can infer more accurate results by summarizing the code-to-comment pattern and achieves 19.58 BLEU score. The IR-based approach and the neural-based approach perform similarly on the whole test set, but their performance differs significantly on these two sets of samples. Thus combining the strengths of these two methods can achieve better performance. 

\begin{table}[t]
\caption{The performance (BLEU) of each approach on the IR-better samples and NMT-better samples}
\centering
\footnotesize
\begin{tabular}{@{}llll@{}}
\toprule
\multirow{2}{*}{Approach} & All   & IR-better samples & NMT-better samples \\ \cmidrule(l){2-4} 
                               & 90908 & 31636 (34.8\%)     & 59272 (65.2\%)      \\ \midrule
\ReCom Retrieve Module    & 18.04 & \textbf{39.55}             & 5.25               \\
DeepCom                  & 20.11 & 20.86             & \textbf{19.58}              \\
\ReCom               & 23.69 & 39.46             & 14.33              \\ \hline \hline
Our Method                     & \textbf{25.45} & 37.5              & 18.0               \\ \bottomrule
\end{tabular}
\label{tab:diss_2part}
\end{table}

By feeding information from the retrieved similar sample (code snippet and comment) to the neural model, the IR+NMT approach, i.e., Re\textsuperscript{2}Com, performs better than the neural-based approach, i.e., DeepCom, on IR-better samples and achieves 39.46 BLEU score. 
However, on NMT-better samples, \ReCom only achieves 14.33 BLEU score, which is 27\% lower than the score of DeepCom. 
{The reason for such a performance degradation is that \ReCom can not accurately distinguish false-positive samples like Figure \ref{fig:IRWorse}, thus incorrectly rely on the inaccurate retrieved information, i.e., the IR-based approach only achieves 5.25 BLEU score on NMT-better samples.} 
Therefore, inaccurate retrieval information can lead to the degradation of the model's generalization. 
In contrast, our approach directly distinguishes whether the retrieved result is accurate, which can help avoid the inaccurate retrieved information misleading the NMT to generate inaccurate comment. 
Thus our approach can outperform \ReCom on the NMT-better samples and the whole test set. 
{Since the Cross-Encoder based classifier cannot perfectly predict whether the IR result is accurate, some samples incorrectly use inaccurate IR results as output or neglect accurate IR results. There is still a distance from the optimal performance of combing IR results and NMT results, i.e., achieving 39.55 BLEU score on IR-better samples and achieving 19.58 BLEU score on NMT-better samples.}


\subsection{Performance of Our Approach on An Alternative Dataset}

{To show the generalization of our approach}, we further verify the performance of our method on another large-scale dataset, i.e., the DeepCom dataset~\cite{hu2020deep}. The DeepCom dataset was collected from GitHub’s Java repositories created from 2015 to 2016 and contained 
445,812 code-comment pairs for training and 20,000 code-comment pairs for validation and testing. 

{We re-run our approach and the three baselines 
on the DeepCom Dataset, and the results are shown in Table \ref{tab:diss_deepcom}.}
Overall, all four methods achieve outstanding performance on the DeepCom dataset, which quite different from their performance on the FunCom dataset.
The main reason can be that the projects used in these two datasets are different, in which more code snippets and comments are reused among projects. 
The IR-based approach, \ReCom retrieval module, achieves 55.28 BLEU score {on the test set}, which implies that code reuse is more frequent on the projects collected by the DeepCom dataset. Thus the neural model can predict the samples in the test set more accurately due to the presence of similar samples in the training set. The neural-based approach, DeepCom, achieves 38.79 BLEU score, which seems to perform well, but it is even inferior to the naive IR-based method. By feeding codes and comments from retrieved similar samples, the IR+NMT method, Re\textsuperscript{2}Com, achieves 50.21 BLEU score on the test set. However, the performance of \ReCom is still worse than the naive IR-based method, which implies that it fails to combine the strengths of the IR-based and NMT-based method on the DeepCom dataset. In contrast, our proposed approach, dynamically combining the generated results from DeepCom and IR-based approach, achieves 57.13 BLEU score on the test set, which successfully combines the strengths of the IR method and NMT method and achieves the best performance.

\begin{table}[t]
\centering
\footnotesize
\caption{The performance of each approach on the DeepCom dataset}
\begin{tabular}{@{}llllll@{}}
\toprule
Approach    & BLEU           & BLEU\textsubscript{1}         & BLEU\textsubscript{2}        & BLEU\textsubscript{3}         & BLEU\textsubscript{4}         \\ \midrule
DeepCom                 & 38.79          & 54.9           & 38.75         & 33.78          & 31.5           \\
\ReCom                  & 50.21          & 61.83          & 50.6          & 46.29          & 43.89          \\
\ReCom Retrieval Module & 55.28          & 65.93          & 55.27         & 51.69          & 49.59          \\ \hline \hline
Our  Method & \textbf{57.13} & \textbf{68.91} & \textbf{57.2} & \textbf{53.07} & \textbf{50.92} \\ \bottomrule
\end{tabular}
\label{tab:diss_deepcom}
\end{table}

\subsection{Effort Saved Comparing to The Existing Combination}


Compared to the existing combination of IR- and NMT-based comment generation approaches, which use both the two models to generate a comment for each input sample, our approach dynamically selects the model to be used. 
To show the effort our method can save, we count the number of samples that do not need to run neural-based approaches to generate comments.

Specifically, our Cross-Encoder based classifier identifies 18,912 samples and 12,979 samples on the FunCom dataset and DeepCom dataset, respectively, that can be directly used for IR results. It implies that about 20\% and 65\% of the samples do not need to be fed into the NMT.
Our approach can save the redundant effort of NMT predicting, making it faster than the current IR+NMT approach. 

\subsection{Threats to Validity}
\label{sec:threats}
\textbf{Internal Validity} relates to the errors in the implementation of the baselines. To mitigate this issue, we directly use the public available code of DeepCom~\cite{hu2020deep}, (ast-)attendgru~\cite{leclair2019neural}, \ReCom~\cite{wei2020retrieve}, and Rencos~\cite{zhang2020retrieval} to implement baselines. Our experiments showed these baselines achieve comparable performance with the result reported in their papers.

\textbf{External Validity} is about the quality of our dataset. Different data sources
{can have significant different characterics}. Therefore, both our proposed approach and the baselines may perform differently on different datasets. In this paper, we only evaluate our proposed approach and baselines on two widely used datasets, i.e., DeepCom~\cite{hu2020deep} and FunCom~\cite{leclair2019neural}. 
In our future work, we will experiment with other datasets.

\textbf{Construct  Validity} relates to the suitability of our evaluation metrics. We use BLEU, ROUGE-L, METEOR, and CIDER to evaluate the generated comments of our approach and other baselines. These metrics mainly measure the gap between generated comments and ground truth in terms of textual similarity. 
\section{Related Work}
\label{sec:rw}
\textbf{Comment generation.} Code comment generation techniques can be divided into three types: manually-crafted templates~\cite{sridhara2010towards,moreno2013automatic}, IR-based~\cite{haiduc2010supporting,haiduc2010use,eddy2013evaluating,wong2015clocom,edmund2014mining}, and neural models~\cite{iyer2016summarizing,hu2018deep,hu2020deep,leclair2019neural,zhang2020retrieval,wei2020retrieve}.

Early studies leveraged manually-craft templates to generate comments automatically. Sridhara et al.~\cite{sridhara2010towards} built the Software Word Usage Model (SWUM) to capture the meaning and relationship of terms in the source code, then organized them into readable comments using different predefined templates. Moreno et al.~\cite{moreno2013automatic} used heuristic rules to capture critical information from the source code and further used them to generate comments. 

Information retrieval (IR) techniques are also widely used in comment generation. One way is to provide extractive summaries of the source code, using IR techniques to extract keywords from the source code and compose them into term-based comments. Haiduc et al.~\cite{haiduc2010supporting,haiduc2010use} treated each function of source code as a document and leveraged Vector Space Model and Latent Semantic Indexing (LSI) to extract relevant terms from source code, then organized selected terms into comments. Eddy et al.~\cite{eddy2013evaluating} took a similar idea and adopted a hierarchical topic model for improvement. Another way is directly use the existing comment of a similar sample. Since code reuse and cloning are common in software development, similar code snippets that use the same code fragments may be found in large project repositories (e.g., GitHub) or software Q\&A sites (e.g., Stack Overflow). Edmund et al.~\cite{wong2015clocom,edmund2014mining} retrieved the replicated samples from the corpus by code clone detection techniques. 

More and more researchers have focused on neural-based methods, which train probabilistic models from large-scale source code in recent years. Iyer et al.~\cite{iyer2016summarizing} treated code to comment as an end-to-end translation problem and first introduced neural machine translation (NMT) into comment generation. They leveraged an attentional seq2seq model to translate code to comment, which used token embedding as the encoder and an LSTM layer as the decoder. Other researchers followed this way. Hu et al.~\cite{hu2018deep} argued that treating code as natural language sequences may lose its syntactical information. They proposed a new structure-based traversal (SBT) method to flatten the AST into sequence and replaced code with it as the model input. Later they proposed another hybrid model~\cite{hu2020deep} that simultaneously used codes and AST sequences for prediction. LeClair et al.~\cite{leclair2019neural} also proposed a similar hybrid model but proved that the neural model also works with only the AST sequence known. The NMT-based method can automatically learn code to comment patterns from the corpus, which saves the manual effort to design features or templates and brings impressive generalization capability. The IR-based method may fail when there are no similar samples in the training set, but the NMT-based method can give more accurate answers.

\textbf{IR-based Neural Comment Generation.} The neural models are difficult to generate low-frequency tokens~\cite{koehn2017rareword}. LeClair et al.~\cite{leclair2019neural} showed that about 21\% of comments in their test set contained low-frequency words (frequency $\leq$100). However, only 7\% generated results of their method contained low-frequency words. The IR-based methods leverage existing comments from similar samples, which may contain low-frequency words and project-specific information. Therefore, researchers have begun to combine IR-based methods with NMT-based methods by feeding information from similar samples (their codes only/ and comments) to assist neural models in better generating low-frequency words. Zhang et al.~\cite{zhang2020retrieval} proposed an approach that fuzed decoded results of the input code snippet and its similar code snippets, which were retrieved based on syntactic similarity and semantical similarity. Wei et al.~\cite{wei2020retrieve} treated the existing comments of similar codes as exemplars, which can be reference examples for generating new comments. They introduced additional encoders to encode codes and comments from similar samples, then jointly trained model. To avoid the disturbance of inaccurate search results, both models decided the degree of using retrieved information based on the embedding similarity of the input and retrieved code snippets. The result shows that these methods can improve both the performance of generated comments and generating low-frequency words. However, both methods may be confused by false-positive samples like Figure \ref{fig:IRWorse}. Without supervised learning, the input and retrieved code snippet of this example will yield similar embedding, making the model mistakenly believe that the retrieved results are accurate and wrongly rely on the inaccurate retrieved result,
{and leading to a decrease in generalization performance. In our work, we treat determining whether the retrieved result is accurate as a supervision task to distinguish false-positive retrieval results more accurately, and combine the IR-based and NMT-based methods in a dynamic manner to avoid the neural model over-rely on the retrieved information.}

\section{Conclusion}
\label{sec:con}
In this paper, we propose a dynamic approach to combine the strength of the IR-based and neural-based comment generation approaches.  
{Specifically, given an input code snippet, we first use an IR-based technique to retrieve a similar code snippet from the corpus. Then we use a Cross-Encoder based classifier to decide the comment generation method to be used dynamically, i.e., if the retrieve similar code snippet is a true positive, we directly use the comment generated by IR-based approach. Otherwise, we input it to the neural-based approach to generate its comment.}  
We have evaluated the effectiveness and generality of our approach on a  large-scale Java dataset. The results show that our approach outperforms the state-of-the-art baselines by a significant margin. 

\bibliographystyle{ACM-Reference-Format}
\bibliography{reference}

\end{document}